\documentclass[12pt]{article}
\usepackage{amssymb}
\usepackage{epsfig}
\usepackage{float}
\usepackage{slashed}
\newcommand{\be}{\begin{equation}}
\newcommand{\ee}{\end{equation}}
\newcommand{\bea}{\begin{eqnarray}}
\newcommand{\eea}{\end{eqnarray}}

\begin{document}
\begin{flushright}
{hep-ph/yymmnnn}
\end{flushright}
\begin{flushright}
CERN-TH/2007-150
\end{flushright}

\vskip 1cm

\begin{center}
 \textbf{TOWARDS A MAXIMAL MASS MODEL}

\end{center}

\begin{center}
V. G. Kadyshevsky
\end{center}
\begin{center}
{\em  Joint Institute
for Nuclear Research, Dubna, R-141980, Russia, and\\
CERN, PH-TH, 1211 Geneva, Switzerland.}
\end{center}

\begin{center}
 M. D. Mateev
\end{center}
\begin{center}
{\em Faculty of Physics, University of Sofia "St. Kliment Ohridsky", 1164
Sofia, Bulgaria, and CERN, PH-TH, 1211 Geneva, Switzerland.}
\end{center}
\begin{center}
V. N. Rodionov
\end{center}

\begin{center}
{\em  Moscow State Geological Prospecting University, 118617 Moscow,
Russia.}
\end{center}
\begin{center}
A. S. Sorin
\end{center}

\begin{center}
{\em  Joint Institute for Nuclear Research, Dubna, R-141980, Russia.}
\end{center}
\abstract{We investigate the possibility to construct a generalization of
the Standard Model, which we call the Maximal Mass Model because it
contains a limiting mass $M$ for its fundamental constituents.  The
parameter $M$ is considered as a new universal physical constant of Nature
and therefore is called the fundamental mass. It is introduced in a purely
geometrical way, like the velocity of light  as a maximal velocity in the
special relativity. If one chooses the Euclidean formulation of quantum
field theory, the adequate realization of the limiting mass hypothesis is
reduced to the choice of the de Sitter geometry as the geometry of the
4-momentum space. All fields, defined in de Sitter p-space in
configurational space obey five dimensional Klein-Gordon type equation
with fundamental mass $M$ as  a mass parameter. The role of dynamical
field variables is played by the Cauchy initial conditions given at $x_5 =
0$, guarantying the locality and gauge invariance principles.  The
corresponding to the geometrical requirements formulation of the theory of
scalar, vector and spinor fields is considered in some detail. On a simple
example it is demonstrated that the spontaneously symmetry breaking
mechanism leads to renormalization of the fundamental mass $M$. A new
geometrical concept of the chirality of the fermion fields is introduced.
It would be responsible for new measurable effects at high energies $E
\geq M$. Interaction terms of a new type, due to the existence of the
Higgs boson are revealed. The most intriguing prediction of the new
approach is the possible existence of exotic fermions with no analogues in
the SM, which  may be candidate for dark matter constituents.}

\section{Introductory remarks}

For decades we have witnessed the impressive success of the Standard Model
(SM) in explaining properties and regularities observed in  experiments
with elementary particles. The mathematical basis of the SM is local
lagrangian quantum field theory (QFT). The very concept of elementary
particle assumes that it does not have a composite structure. In agreement
with the contemporary experimental data such a structure has not been
disclosed for no one of the fundamental particles of the SM, up to
distances of the order of $10^{-16} - 10^{-17}$ cm. The adequate
mathematical images of point like particles are the local quantized fields
- boson and spinor. Particles are the quanta of the corresponding fields.
In the framework of the SM these are leptons, quarks, vector bosons and
the Higgs scalar, all characterized by certain values of mass, spin,
electric charge, colour, isotopic spin, hypercharge, etc.

Intuitively it is clear that the elementary particle should carry small
enough portions of different "charges" and "spins". In the theory this is
guaranteed by assigning the local fields to the lowest  representations of
the corresponding groups.

As for the mass of the particle $m$, this quantity is the Casimir operator
of the \emph{\textbf{noncompact}}  Poincar\'{e} group and in the unitary
representations of this group, used in QFT, they may have arbitrary values
in the interval $0 \leq m <\infty$. In the SM one observe a great variety
in the mass values. For example, t-quark is more than 300000 times heavier
than the electron. In this situation  the question naturally arises: up to
what values of mass one may apply the concept of a local quantum field?
Formally the contemporary QFT remains logically perfect scheme and its
mathematical structure does not change at all up to arbitrary large values
of quanta's masses. For instance, the free Klein-Gordon equation for the
one component real scalar field $\varphi(x)$ has always the form:
\begin{equation}\label{KG}
    (\square + m^2)\varphi(x) = 0.
\end{equation}
From here after standard Fourier transform:
\begin{equation}\label{FT}
    \varphi(x) = \frac{1}{(2\pi)^{3/2}} \int e^{-i p_\mu x^\mu}\;\varphi(p)\;d\,^4
    p\;\;\;\;\; ( p_\mu x^\mu = p^0 x^0 - \emph{\textbf{p.}} \emph{\textbf{x}} )
\end{equation}
we find the equation of motion in Minkowski momentum 4-space:
\begin{equation}\label{KGp}
    ( m^2 - p^2 ) \varphi(p) = 0, \;\;\;\;\;\; p^2 = p_0^2 -
    \emph{\textbf{p}}^2.
\end{equation}
From geometrical point of view $m$ is the radius of the "mass shell"
hyperboloid:
\begin{equation}\label{ms}
    m^2 = p_0^2 -
    \emph{\textbf{p}}^2,
\end{equation}
where the field $\varphi(p)$ is defined and in the Minkowski momentum
space one may embed hyperboloids of the type (\ref{ms}) of arbitrary
radius.

In 1965 M. A. Markov  \cite{Markov} pioneered the hypotheses according to
which the mass spectrum of the elementary particles should be cut off at
the Planck mass $m_{Planck} = 10^{19} GeV $ :
\begin{equation}\label{mpl}
 m \leq m_{Planck}.
\end{equation}
The particles with the limiting mass $m = m_{Planck}$, named by the author
"maximons"  should  play special role in the world of elementary
particles. However, Markov's original condition (\ref{mpl}) was purely
phenomenological and he used standard field theoretical techniques even
for describing the maximon.

In \cite{Kad} - \cite{KMRS} a more radical approach was developed. The
Markov' s idea about existence of a maximal value for the masses of the
elementary particles has been understood as a new fundamental principle of
Nature, which similarly to the relativistic and quantum postulates should
be put in the grounds of QFT. Doing this the condition of finiteness of
the mass spectrum should be introduced by the relation:
\begin{equation}\label{Mfund}
m \leq  M,
\end{equation}
where the maximal mass parameter M called the "\emph{\textbf{fundamental
mass}}" is a \emph{\textbf{new universal physical constant}}.

A \emph{\textbf{new concept}} of a local quantum field has been developed
on the ground of (\ref{Mfund}) an on simple geometric arguments, the
corresponding Lagrangians were constructed  and an adequate formulation of
the principle of local gauge invariance has been found. It has been also
demonstrated that the fundamental mass $M$ in the new approach plays the
role of an independent universal scale in the region of ultra high
energies $E \geq M $.

It is worth emphasizing that here, due to eq(\ref{Mfund}), the Compton
wave length of a particle $\lambda_C = \hbar/ m c$ can not be smaller than
the "\emph{\textbf{fundamental length}}" $l = \hbar/ M c$. According to
Newton an Wigner \cite {NW} the parameter $\lambda_C$ characterizes the
dimensions of the region of space in which a relativistic particle of mass
$m$ can be localized. Therefore the fundamental length $l$ introduces into
the theory an universal bound on the accuracy of the localization in space
of elementary particles.

The objective of the present work, in few words, is to include the
principle of maximal mass (\ref{Mfund}) into the basic principles of the
Standard Model. The appearing in this way new scheme, which we called
Maximal Mass Model, from our point of view is interesting already because
in it are organically bound the trusted methods of the local gauge QFT and
elegant, even not as popular, geometric ideas.

The paper is organized as follows. In the next section  on the example of
neutral scalar field it is demonstrated how on can use geometrical
arguments to construct a version of a QFT internally consistent with the
principle of a maximal mass (\ref{Mfund}). On the same example we describe
the simplest mechanism of spontaneous breaking of discrete symmetry.
Series of relations from the new formulation of the vector field theory
are given.

Section 3. is dedicated to the description of the fermion fields in the
new approach. A special attention is given to the new geometrical
definition of \emph{\textbf{chirality}}, which may be used at all values
of energy, including the region $E \geq M$. The most intriguing prediction
of the new approach implies that a new family of "exotic" fermions with
non standard properties of polarizations have to exist in Nature. These
earlier unknown Fermi fields, as we hypothesize, are the constituents of
"dark matter".

In Section 4. some questions in relation to the description of the
electro-weak interactions in the new version of the SM are considered and
some high energy effects, whose magnitude depends on the fundamental mass
$M$, are predicted.

\section{Boson fields in de Sitter momentum space.}

Let us go back to the free one component real scalar field we considered
above (\ref {KG} - \ref{KGp}). We shall suppose that its mass $m$
satisfies the condition (\ref{Mfund}). How  should one modify the
equations of motion in order that the existence of the bound (\ref{Mfund})
should become as evident as it is the limitation $v \leq c$ in the special
theory of relativity? In the latter case everything is explained in a
simple way: the relativization of the 3-dimensional velocity space is
equivalent to transition in this space from Euclidean to Lobachevsky
geometry, realized on the 4-dimensional hyperboloid \footnote{ To be exact
on the upper sheet of this hyperboloid.}(\ref{ms}).  Let us act in  a
similar way and change the 4-dimensional Minkowski momentum space, which
is used in the standard QFT, to anti de Sitter momentum space, realized on
the 5-hyperboloid:
\begin{equation}\label{ads}
    p^2_0 - \emph{\textbf{p}}^2 + p_5^2 = M^2.
\end{equation}
We shall suppose that in p-representation our scalar field is defined just
on the surface (\ref{ads}), i.e. it is a function of five variables $(p_0,
\emph{\textbf{p}}, p_5)$, which are connected by the relation (\ref{ads}):
\begin{equation}\label{dads}
    \delta(p^2_0 - \emph{\textbf{p}}^2 + p_5^2 - M^2)\varphi(p_0, \emph{\textbf{p}},
    p_5).
\end{equation}
The energy $p_0$ and the 3-momentum $\emph{\textbf{p}}$ here preserve
their usual sense and the mass shell relation (\ref{ms}) is satisfied as
well. Therefore,  for the considered field $\varphi(p_0,
\emph{\textbf{p}}, p_5)$ the condition (\ref{Mfund}) is always fulfilled.

Clearly in eq. (\ref{dads}) the specification of a single function
$\varphi(p_0, \emph{\textbf{p}}, p_5)$ of five variables $(p_\mu, p_5)$ is
equivalent to the definition of two independent functions $\varphi_1(p)$
and $\varphi_2(p)$ of the 4-momentum $p_\mu$:
\begin{equation}\label{12}
    \varphi(p_0, \emph{\textbf{p}}, p_5) \equiv \varphi(p, p_5) = \left(%
\begin{array}{c}
  \varphi(p, p_5) \\
  \varphi(p, - p_5) \\
\end{array}%
\right) = \left(%
\begin{array}{c}
  \varphi_1(p) \\
  \varphi_2(p) \\
\end{array}%
\right), |p_5| = \sqrt{M^2 - p^2}.
\end{equation}

The appearance of the new discrete degree of freedom $p_5/|p_5|$ and the
associated doubling of the number of field variables is a most important
feature of the new approach. It must be taken into account in the search
of the equation of motion for the free field in de Sitter momentum space.
Because of the mass shell relation (\ref{ms}) the Klein - Gordon equation
(\ref{KGp}) should be also satisfied by the field $\varphi(p_0,
\emph{\textbf{p}}, p_5)$ :
\begin{equation}\label{KGp2}
    ( m^2 - p_0^2 + \emph{\textbf{p}}^2)  \varphi(p_0, \emph{\textbf{p}}, p_5) =
    0.
\end{equation}
From our point of view this relation is unsatisfactory for 2 reasons:

1. It does not reflect the bounded mass condition (\ref{Mfund}).

2. It can not be used to determine the dependence of the  field on the new
quantum number $p_5/|p_5|$ in order to distinguish between the components
$\varphi_1(p)$ and  $\varphi_2(p)$.

Here we notice that, because of (\ref{ads}) eq.(\ref{KGp2}) may be written
as:
\begin{equation}\label{razlKG}
    (p_5 + M \cos \mu)(p_5 - M \cos \mu)\varphi(p, p_5) = 0 , \;\;\;\;\; \cos
    \mu = \sqrt{1 - \frac{m^2}{M^2}}.
\end{equation}
Now, following the  Dirac trick we postulate the  equation of motion under
question in the form:
\begin{equation}\label{NKG}
    2M(p_5 - M \cos \mu)\varphi(p, p_5) = 0 .
\end{equation}
Clearly, eq. (\ref{NKG}) has none of the enumerated defects present in the
standard Klein-Gordon equation (\ref{KGp}). However, equation (\ref{KGp})
is still  satisfied by the field $\varphi(p, p_5)$.

From eqs. (\ref{NKG}) and (\ref{12}) it follows that:
\begin{equation}\label{NKG12}
    \begin{array}{c}
      2M(|p_5| - M \cos \mu)\varphi_1(p) = 0, \\ \\
      2M(|p_5| + M \cos \mu)\varphi_2(p) = 0, \\
    \end{array}
\end{equation}
and we obtain:
\begin{equation}\label{122}
    \begin{array}{c}
      \varphi_1(p) = \delta(p^2 - m^2)\widetilde{\varphi}_1(p)\\ \\
      \varphi_2(p) = 0\\
    \end{array}
\end{equation}
Therefore, the free field $\varphi(p, p_5)$ defined in anti de Sitter
momentum space (\ref{ads}) describes the same free scalar particles of
mass $m$ as the field $\varphi(p)$ in Minkowski p-space, with the only
difference that now we necessarily have $m \leq M$. The two component
structure (\ref{12}) of the new field does not manifest itself on the mass
shell, owing to (\ref{122}). However, it will play an important role when
the fields interact  - i.e off the mass shell.

Now we face the problem of constructing the action corresponding to eq.
(\ref{NKG}) and transforming it to configuration representation.

Due to mainly technical reasons \footnote{The corresponding comments on
the topic will be given a bit later.}  in the following we shall use the
Euclidean formulation of the theory, which appears as an analytical
continuation to purely imaginary energies:
\begin{equation}\label{eup}
    p_0 \rightarrow ip_4.
\end{equation}
In this case instead of the anti de Sitter p-space (\ref{ads}) we shall
work with de Sitter p-space:
\begin{equation}\label{ds}
- p_n^2 + p_5^2 = M^2, \;\;\;\;\;\; n = 1,2, 3, 4.
\end{equation}
Obviously:
\begin{equation}\label{p5}
    p_5 = \pm\sqrt{M^2 + p^2}.
\end{equation}
If one uses eq. (\ref{ds}), the Euclidean Klein-Gordon operator $m^2 +
p^2$ may be written, similarly to (\ref{razlKG}) in the following
factorized form:
\begin{equation}\label{factf}
m^2 + p^2 = (p_5 + M \cos \mu)(p_5 - M \cos \mu).
\end{equation}
Clearly the non-negative functional:
\begin{equation}\label{action}
\begin{array}{c}
  S_0(M) = \pi M \times \\ \\
  \int \frac{d^4 p }{|p_5|}\left[\varphi^+_1(p)2M (|p_5| - M \cos
    \mu)\varphi_1(p) + \varphi^+_2(p)2M (|p_5| + M \cos
    \mu)\varphi_2(p)\right], \\
\end{array}
\end{equation}
\begin{equation}\label{123}
    \varphi_{1,2}(p) \equiv \varphi(p, \pm |p_5|),
\end{equation}
plays the role of the action integral of the free Euclidean field
$\varphi(p, p_5)$. The action may be  written also as a 5 - integral:
\begin{equation}\label{5act}
    \begin{array}{c}
       S_0(M) = 2\pi M \times  \\ \\
       \int \varepsilon(p_5) \delta(p_L p^L - M^2)d^5 p \left[\varphi^+(p, p_5)2M (p_5 - M \cos
    \mu)\varphi(p, p_5) \right],\\ \\
     L = 1, 2, 3, 4, 5,
\end{array}
\end{equation}
where
\begin{equation}\label{epsilon}
    \varepsilon(p_5) = \frac{p_5 }{|p_5|}.
\end{equation}
 The Fourier transform and the configuration representation have special
 role in this approach. First, we note that in the basic equation (\ref{ds})
 which defines de Sitter p-space, all the components of the 5-momentum
 enter on equal footing. Therefore the expression $\delta(p_L p^L - M^2)\varphi(p,
 p_5)$, which now replaces (\ref{dads}) may be Fourier transformed:

 \begin{equation}\label{FT2}
    \frac{2M}{(2\pi)^{3/2}}\int e^{-i p_K x^K}\delta(p_L p^L - M^2)\varphi(p,
 p_5) d^5\; p = \varphi(x, x_5), \;\; K, L = 1, 2, 3, 4, 5.
\end{equation}
This function obviously satisfies the following differential equation in
\emph{\textbf{5-dimensional configuration space}}:
\begin{equation}\label{pent}
    \left(\frac{\partial^2}{\partial x_5^2 } - \square + M^2\right)\varphi (x, x_5) =
    0.
\end{equation}
Integration over $p_5$ in (\ref{FT2}) gives:
\begin{equation}\label{FT3}
\begin{array}{c}
  \varphi(x, x_5) = \frac{2M}{(2\pi)^{3/2}}\int e^{i p_n x^n} \frac{d^4
p}{|p_5|}
  \left[e^{-i |p_5| x^5}\varphi_1(p) + e^{i |p_5| x^5}\varphi_2(p)\right],
  \\ \\
 \varphi^+(x, x_5) = \varphi(x, - x_5),  \\
\end{array}
\end{equation}
from which we get:
\begin{equation}\label{FT4}
\frac{i}{M}\frac{\partial\varphi(x, x_5)}{\partial x_5} =
\frac{1}{(2\pi)^{3/2}}\int e^{i p_n x^n} d^4 p
  \left[e^{-i |p_5| x^5}\varphi_1(p) - e^{i |p_5| x^5}\varphi_2(p)\right],
\end{equation}

The four dimensional integrals (\ref{FT3}) and (\ref{FT4})
 transform the fields $\varphi_1(p)$ and $\varphi_2(p)$ to the configuration representation.
 The inverse transforms have the form:
 \begin{equation}\label{IFT}
    \begin{array}{c}
      \varphi_1(p) = \frac{-i}{2M (2\pi)^{5/2}} \int e^{-i p_n x^n} d^4 x \left[\varphi(x, x_5)
      \frac{\partial e^{i |p_5|
      x^5}}{\partial x_5} -
      e^{i |p_5| x^5}\frac{\partial\varphi(x, x_5)}{\partial x_5}\right],\\ \\
      \varphi_2(p) = \frac{i}{2M (2\pi)^{5/2}} \int e^{-i p_n x^n} d^4 x \left[\varphi(x, x_5)
      \frac{\partial e^{- i |p_5|
      x^5}}{\partial x_5} -
      e^{ - i |p_5| x^5}\frac{\partial\varphi(x, x_5)}{\partial x_5}\right].  \\
    \end{array}
\end{equation}

We note that the independent field variables:
\begin{equation}\label{phix}
    \varphi(x, 0)\equiv \varphi(x) = \frac{2M}{(2\pi)^{3/2}}\int e^{i p_n x^n} d^4\,p
  \frac{\varphi_1(p) + \varphi_2(p)}{|p_5|}
\end{equation}
and
\begin{equation}\label{chix}
    \frac{i}{M}\frac{\partial\varphi(x, 0)}{\partial x_5} \equiv \chi(x)=
\frac{1}{(2\pi)^{3/2}}\int e^{i p_n x^n} d^4 p
  \left[\varphi_1(p) - \varphi_2(p)\right]
\end{equation}
can be treated as  initial Cauchy data on the surface $x_5 = 0$ for the
 hyperbolic-type equation (\ref{pent}).

 Now substituting eq.(\ref{IFT}) into the action (\ref{action}) we obtain:
 \begin{equation}\label{action2}
 \begin{array}{c}
  S_0(M) =  \frac{1}{2}\int d^4 \,x \left[\left|\frac{\partial \varphi(x, x_5)}{\partial x_n}\right|^2 +
    m^2|\varphi(x, x_5)|^2 + \left|i \frac{\partial \varphi(x, x_5)}{\partial x_5} - M \cos
    \mu \varphi(x, x_5)\right|^2\right] \equiv \\ \\
   \equiv \int L_0 (x, x_5)d^4 \,x .\\
 \end{array}
\end{equation}
It is easily verified that due to eq. (\ref{pent}) the action
(\ref{action2}) is independent of $x_5$:
\begin{equation}\label{ds0dx5}
    \frac{\partial S_0(M)}{\partial x_5} = 0.
\end{equation}
Therefore the variable $x_5$ may be arbitrarily fixed and $S_0(M)$ may be
viewed as a functional of the corresponding initial Cauchy data for the
equation (\ref{pent}). For example, for $x_5 =0$ we have:
\begin{equation}\label{S0M}
\begin{array}{c}
    S_0(M) =  \frac{1}{2}\int d^4 \,x \left[\left(\frac{\partial \varphi(x)}{\partial x_n}\right)^2 +
    m^2(\varphi(x))^2 + M \left(\chi(x) - \cos
    \mu \varphi(x)\right)^2\right] \equiv \\ \\
   \equiv \int L_0 (x, M)d^4 \,x .
   \end{array}
\end{equation}
We have thus  shown that in the developed  approach  the property of
locality of the theory  does not disappear, moreover it becomes even
deeper, as it is extended to dependence on the extra fifth dimension
$x_5$.

The new Lagrangian density $L_0 (x, x_5)$ [see (\ref{action2})] is a
Hermitian form constructed from $\varphi(x, x_5)$ and the components of
the 5-component gradient $\frac{\partial \varphi(x)}{\partial x_L}, (L =
1, 2, 3, 4, 5).$ It is clear that although $L_0 (x, x_5)$ depends
explicitly on $x_5$, the theory essentially remains
\emph{\textbf{four-dimensional}} [see eq. (\ref{ds0dx5}) and (\ref{S0M})].

As may be seen from the transformations which have been made, the
dependence of the action (\ref{S0M}) on the two functional arguments
$\varphi(x)$ and $\chi(x)$ is a direct consequence of the fact that in
momentum space the field has a doublet structure $\left(\begin{array}{c}
                                                  \varphi_1(p) \\
                                                    \varphi_2(p)
                                                  \end{array}\right)$ due
                                                  to the two possible
                                                  values of $p_5$.
However, the Lagrangian $L_0 (x, M)$ does not contain a kinetic term
corresponding
 to the field $\chi(x)$. Therefore, this variable is just auxiliary.

 The special role of the 5-dimensional configuration  space in the new formalism is determined
 by the fact that the gauge symmetry transformations are localized
now in it. The initial data for the  equation (\ref{pent}):
\begin{equation}\label{id}
    \left(%
\begin{array}{c}
  \varphi(x, x_5) \\ \\
  \frac{i}{M} \frac{\partial\varphi(x, x_5)}{\partial x_5} \\
\end{array}%
\right)_{x_5 = fixed\;\; value}
\end{equation}
 are
subject to these transformations.

Let us now discuss this point in more detail, supposing that the field
$\varphi(x, x_5)$ is not Hermitian and some internal symmetry group is
associated with it:
\begin{equation}\label{isg}
    \varphi' = U \varphi.
\end{equation}
Upon localization of the group in the 5-dimensional x-space:
\begin{equation}\label{lisg}
    U \rightarrow U(x, x_5),
\end{equation}
the following gauge transformation law arises for the initial data
(\ref{id}) on the plane $x_5 = 0$:
\begin{equation}\label{gt}
    \begin{array}{c}
      \varphi'(x) = U(x, 0)\varphi(x), \\ \\
      \chi'(x) = \frac{i}{M} \frac{\partial U(x, 0)}{\partial x_5}\varphi(x) + U(x, 0)\chi(x).\\
    \end{array}
\end{equation}
The group character of the transformations (\ref{gt}) is obvious. The
specific form of the matrix $U(x, x_5)$ can be determined in the new
theory of vector fields, which is a generalization of the standard theory
in the spirit of our approach ( see the end of this section).

It is clear that the equation (\ref{pent}) may be represented as a system
of two equations of first order in the derivative
$\frac{\partial}{\partial x_5}$ \cite{KadFursBos}:
\begin{equation}\label{fode}
    \left\{ \frac{i}{M} \frac{\partial }{\partial x_5} - \left[\sigma_3 \left(1 - \frac{\square}{2
    M^2}\right)-i\sigma_2 \frac{\square}{2
    M^2}
    \right]\right\} \phi(x. x_5) = 0,
\end{equation}
where
\begin{equation}\label{phixx5}
    \phi(x, x_5) = \left(%
\begin{array}{c}
  \frac{1}{2} \left[\varphi(x, x_5) +  \frac{i}{M}\frac{\partial\varphi(x, x_5) }{\partial
  x_5}\right]\\ \\
  \frac{1}{2} \left[\varphi(x, x_5) -  \frac{i}{M}\frac{\partial\varphi(x, x_5) }{\partial x_5}\right]\\
\end{array}%
\right) \equiv \left(%
\begin{array}{c}
  \phi_I(x, x_5) \\ \\
  \phi_{II}(x, x_5)\\
\end{array}%
\right),
\end{equation}
($\sigma_i, i = 1, 2, 3$  are the Pauli matrices). If we compare
(\ref{phixx5}) with (\ref{phix}) and (\ref{chix}) we find relations
between the initial Cauchy data for the  equation (\ref{pent}) and the
system (\ref{fode}):
\begin{equation}\label{con}
    \phi(x, 0) = \left(%
\begin{array}{c}
  \phi_I(x, 0) \\ \\
  \phi_{II}(x, 0)\\
\end{array}%
\right) = \left(%
\begin{array}{c}
 \frac{1}{2}(\varphi(x) + \chi(x)) \\ \\
 \frac{1}{2}(\varphi(x) - \chi(x))  \\
\end{array}%
\right) \equiv \phi(x).
\end{equation}
It easy to show that in the basis (\ref{con}) the Lagrangian $L_0 (x, M)$
from (\ref{S0M}) looks like the following:
\begin{equation}\label{L0MI}
L_0 (x, M) = \frac{\partial\phi(x) }{\partial
  x_n}(1 + \sigma_1)\frac{\partial\phi(x) }{\partial
  x_n} + 2 M^2\phi(x)(1 - \cos\mu\,\sigma_3)\phi(x).
\end{equation}

Let us discuss now the question about the conditions of the transition of
the new scheme into the standard Euclidean QFT (the so called
"correspondence principle"). The Euclidean momentum 4-space is the "flat
limit" of de Sitter p-space and may be associated with the approximation:
\begin{equation}\label{FlatLp}
    \begin{array}{c}
      |p_n| \ll M \\
      p_5 \simeq M \\
    \end{array}
\end{equation}
In the same limit in configuration space we shall have:
\begin{equation}\label{Flatlx}
\begin{array}{c}
  \varphi(x, x_5) = e^{-iM x_5}\varphi(x)  \\
  \chi(x) = \varphi(x) \\
\end{array}
\end{equation}
or
\begin{equation}\label{Flatlx1}
   \phi(x) = \left(%
\begin{array}{c}
  \varphi(x) \\
  0 \\
\end{array}%
\right)
\end{equation}
With the help of (\ref{fode}) it is not difficult to obtain \cite{KadIbad,
kadFursSpin} the  corrections of the order of $O(\frac{1}{M^2})$ to the
zero approximation (\ref{Flatlx1}):
\begin{equation}\label{1cor}
     \phi(x) = \left(%
\begin{array}{c}
  \left(1 - \frac{\square}{4 M^2}\right)\varphi(x) \\ \\
  \frac{\square}{4 M^2} \varphi(x) \\
\end{array}%
\right)
\end{equation}
from which ( see eq. (\ref{con})) we have:
\begin{equation}\label{phi-chifl}
\varphi(x) - \chi(x) = \frac{\square \varphi(x)}{2 M^2}
\end{equation}
Taking into account (\ref{phi-chifl}) and (\ref{razlKG}) one may conclude
that in the "flat limit" (formally when $M \rightarrow \infty$) the
Lagrangian $L_0(x, M)$ from (\ref{S0M}) coincides with its Euclidean
counterpart.

A key role in the SM belongs to the scalar Higgs field, the interactions
with which allows the other fields to get masses. As far as in our model
the masses of all particles, including the mass of the Higgs boson itself,
should obey the condition (\ref{Mfund}), one would presume that there
exists a deep internal connection between the Higgs field and the
fundamental mass $M$. As a matter of fact, before the Higgs mechanism is
switched on, all fields by definition are massless \footnote{Higgs boson,
as known,   at this stage is with  mass of a tachyon.} and because of that
the bound (\ref{Mfund}) at this stage has no physical content. Only,
together with the appearance of the mass spectrum of the particles the
condition (\ref{Mfund}) obtains sense and therefore the
 magnitude of $M$  should be essentially fixed by the same Higgs
mechanism.

In order to get some orientation in this situation let us consider, in the
framework of our approach, the  example of the simplest  mechanism,
connected with the spontaneous breaking of a discrete symmetry. In the
beginning, in order to describe the scalar field, let us use the doublet
(\ref{con}). The total Lagrangian $L_{tot}(x)$, in analogy with the
traditional approach, will include a free part (\ref{L0MI}) at $\mu = 0$
and the well known interaction Lagrangian:
\begin{equation}\label{Lint}
    L_{int}(x) = \frac{\lambda^2}{4}(\phi^2 - v^2)^2.
\end{equation}
Therefore, we have:
\begin{equation}\label{Ltot}
    L_{tot}(x) = \frac{\partial\phi(x) }{\partial
  x_n}(1 + \sigma_1)\frac{\partial\phi(x) }{\partial
  x_n} + 2 M^2\phi(x)(1 - \cos\mu\,\sigma_3)\phi(x) + \frac{\lambda^2}{4}(\phi^2 - v^2)^2.
\end{equation}
Here we used the field $\phi(x)$ only to write the interaction
(\ref{Lint}) in the known symmetric form. Now in (\ref{Ltot}) we may go
back to the variables $\varphi(x)$ and $\chi(x)$ (see (\ref{con})):
\begin{equation}\label{Ltot1}
L_{tot}(x) = \frac{1}{2}\left(\frac{\partial \varphi(x)}{\partial
x_n}\right)^2 +
     \frac{M^2}{2}\left(\varphi(x)-\chi(x)\right)^2
  + \frac{\lambda^2}{4}\left(\frac{\varphi^2(x) + \chi^2(x)}{2} -
    v^2\right)^2
    \end{equation}
    The Lagrangian (\ref{Ltot1}) remains invariant under the
    transformation:
    \begin{equation}\label{dt}
    \begin{array}{c}
      \varphi(x) \rightarrow - \varphi(x)\\
      \chi(x)\rightarrow - \chi(x) \\
    \end{array}
\end{equation}
However, this symmetry is spontaneously broken. The transition to a stable
"vacuum" is realized by the transformations:
\begin{equation}\label{sbv}
    \begin{array}{c}
      \varphi(x) = \varphi'(x) + v\\
      \chi(x)= \chi'(x) + v  \\
    \end{array}
\end{equation}
In the new variables $\varphi'(x)$ and $\chi'(x)$ the quadratic in the
fields part of the Lagrangian (\ref{Ltot1}) takes the form:
\begin{equation}\label{Lq}
\frac{1}{2}\left(\frac{\partial \varphi'(x)}{\partial x_n}\right)^2 +
     \frac{1}{2}(M^2 +\frac{ \lambda^2 v^2}{2})\left(\varphi'^2 (x)+\chi'^2 (x)\right)-
     (M^2 - \frac{\lambda^2 v^2}{2})\varphi'(x)\chi'(x).
\end{equation}
Comparing (\ref{Lq}) and (\ref{S0M}) we may  conclude that:

1. In result of the spontaneous  breaking of the symmetry (\ref{dt}) the
fundamental mass $M$ experiences  renormalization:
\begin{equation}\label{ren}
    M^2 \rightarrow M^2 + \frac{\lambda^2 v^2}{2}
\end{equation}

2. The considered scalar particle acquires mass:
\begin{equation}\label{mh}
    m = \sqrt{2}\lambda v \frac{1}{\sqrt{1 + \frac{\lambda^2 v^2}{2 M^2}}},
\end{equation}
which  satisfies the condition \footnote{Let us note that (\ref{condm}) is
equivalent to the inequality $\left(1 - \frac{\lambda
v}{\sqrt{2}M}\right)^2 \geq 0$.}:
\begin{equation}\label{condm}
    m \leq \sqrt{M^2 + \frac{\lambda^2 v^2}{2 }}.
\end{equation}
Therefore, if we, in advance, take into account the renormalization
(\ref{ren}), due to the Higgs mechanism we may write the Lagrangian
(\ref{Ltot1}) in the form \footnote{In order the Lagrangian (\ref{Ltot})
remains positively definite, it is natural to suppose that $M^2
> \frac{\lambda^2 v^2}{2 }$.}:
\begin{equation}\label{Ltotren}
\begin{array}{c}
L_{tot}(x) = \frac{1}{2}\left(\frac{\partial \varphi(x)}{\partial
x_n}\right)^2 +
     \frac{1}{2}(M^2 - \frac{\lambda^2 v^2}{2})\left(\varphi(x)-\chi(x)\right)^2  +
     \\ \\
     + \frac{\lambda^2}{4}(\frac{\varphi^2(x) + \chi^2(x)}{2} - v^2)^2.
\end{array}
\end{equation}
In this way we shall have instead of (\ref{mh}):
\begin{equation}\label{mh1}
    m = \sqrt{2}\lambda v \sqrt{1 - \frac{\lambda^2 v^2}{2 M^2}} \equiv
    m_0\sqrt{1 - \frac{m_0^2}{4 M^2}}
\end{equation}
The quantity $m_0 = \sqrt{2}\lambda v $ is the maximal value of the mass
of the considered scalar particle. It may be reached only in the "flat
limit" $M \rightarrow \infty$, when the Lagrangian (\ref{Ltotren}) because
of (\ref{Flatlx}) and (\ref{phi-chifl}) takes the usual form:
\begin{equation}\label{Ltotus}
    L_{tot}(x) = \frac{1}{2}\left(\frac{\partial \varphi(x)}{\partial
x_n}\right)^2 + \frac{\lambda^2}{4}\left(\varphi^2(x)  -
    v^2\right)^2.
\end{equation}

The concluding part of this section contents a summary of some results of
the paper \cite{CDKMNCG}, in which the formulation of the gauge fields
theory is developed in de Sitter momentum space (\ref{ds}).

 In the new scheme the
electromagnetic potential, similarly to the momentum, becomes a 5-vector
and in p-representation one may consider expressions like:
\begin{equation}\label{Apotds}
    \delta(p_4^2 + \emph{\textbf{p}}^2 - p_5^2 - M^2)A_L(p, p_5), \;\;\;\;L = 1, 2, 3, 4, 5.
\end{equation}
Its 5-dimensional Fourier transform (compare with(\ref{FT2})) has the
form:
\begin{equation}\label{5FTA}
\begin{array}{c}
 A_L(x, x_5) = \frac{2M}{(2\pi)^{3/2}}\int e^{-i p_N x^N}\delta(p_K p^K - M^2)A_L(p,
    p_5)d ^5\,p,  \\ \\
K, L, N = 1, 2, 3, 4, 5. \\
\end{array}
\end{equation}
It is evident that (\ref{5FTA}) satisfies the equation (\ref{pent}):
\begin{equation}\label{pentA}
     \left(\frac{\partial^2}{\partial x_5^2 } - \square + M^2\right) A_L(x,
     x_5) =
    0.
\end{equation}
The action is given by the integrals (compare with (\ref{5act}) and
(\ref{action2}))
\begin{equation}\label{5SA}
\begin{array}{c}
  S_0(M) = 2\pi M \times\\ \\
  \times\int \varepsilon(p_5) \delta(p_L p^L - M^2)d\,^5p \,2 M (p_5 - M
       )\left|A_n(p, p_5) - \frac{p_n A_5(p, p_5)}{p_5 - M}\right|^2 = \\
       \\
  = \int d^4 x L_0 (x, x_5) = \frac{1}{4}\int d^4 x F^*_{KL}(x, x_5)F^{KL}(x, x_5) +
  \\ \\
 + \frac{1}{2}\int d^4 x \left|\frac{\partial(e^{iM x_5}A_L(x,x_5))}{\partial x_L}
 - 2i M e^{iM x_5}A_5(x,x_5)\right|^2,\\ \\
 n = 1, 2, 3, 4;\;\;\;\;\;K, L = 1, 2, 3, 4, 5, \\
\end{array}
\end{equation}
where the "field strength 5-tensor":
\begin{equation}\label{FKL}
F^{KL}(x, x_5) = \frac{\partial(e^{iM x_5}A_K(x,x_5))}{\partial x_L} -
\frac{\partial(e^{iM x_5}A_L(x,x_5))}{\partial x_K}.
\end{equation}
is introduced. This quantity is obviously expressed in terms  of the
commutator of the 5-dimensional covariant derivatives:
\begin{equation}\label{DL}
   D_L = \frac{\partial}{\partial x^L} - i q e^{iM x_5}A_L(x,x_5),
\end{equation}
where $q$ is the electric charge. It is easy to verify that the integral
(\ref{5SA}) is invariant under gauge transformations of the 5-potential
$A_L(x,x_5)$:
\begin{equation}\label{gt1}
e^{iM x_5}A_L(x,x_5) \rightarrow e^{iM x_5}A_L(x,x_5) - \frac { \partial
(e^{iM x_5}\lambda(x,x_5))}{\partial x^L}
\end{equation}
with the condition:
\begin{equation}\label{pentlambda}
\left(\frac{\partial^2}{\partial x_5^2 } - \square + M^2\right) \lambda(x,
     x_5) =
    0.
\end{equation}
Let us notice that the gauge function $\lambda(x,x_5)$ is defined by two
initial data $\lambda(x) = \lambda(x,0)$ and $\mu(x) = \frac{i}{M} \frac {
\partial \lambda(x,0))}{\partial x^5}$. Therefore, the group (\ref{gt}) is
broader than the standard gauge group. This is due to the fact that in the
transition to the 5-dimensional description there appear additional
superfluous gauge degrees of freedom, subject to removal.

Similarly to its scalar analogue (\ref{action2}), the action (\ref{5SA}),
because of (\ref{pentA}), does not depend on the coordinate $x_5$. For
that reason it may be considered as a functional of the Cauchy data:
\begin{equation}\label{CDA}
    A_L(x, 0) = A_L(x),\;\;\;\;\;\frac{i}{M}\frac{\partial A_L(x,0)}{\partial
     x_5} \equiv X_L(x).
\end{equation}
Let us  notice \cite{CDKMNCG}, that:
\begin{equation}\label{prhc}
A^+_n(x)= A_n(x), \;\;\;\;\;A^+_5(x)= - A_5(x).
\end{equation}
Coming back to the relations (\ref{lisg}) - (\ref{gt}) we may assert
 that in the considered Abelean case, because of (\ref{gt1}) and
(\ref{pentlambda}) we have:
\begin{equation}\label{emgt}
    U(x, x_5) = \exp\left(i e^{iMx_5} q \lambda(x, x_5)\right)
\end{equation}
and therefore:
\begin{equation}\label{ggcsf}
    \begin{array}{c}
      \varphi'(x) = e^{i q \lambda(x)}\varphi(x) \\ \\
      \chi'(x) = e^{i q \lambda(x)}[\chi(x) + iq (\mu(x) - \lambda(x))\varphi(x)]\\
    \end{array}
\end{equation}
The considered technique allows us to formulate in our approach an
\emph{\textbf{unique}} prescription for construction of the action
integral of the Euclidean scalar electrodynamics consistent with the
requirements of locality, gauge invariance and de Sitter structure of
momentum space:

1. In the action integral for the complex scalar field (use (\ref{S0M}))
as a pattern) it is necessary to make the \emph{\textbf{minimal
interaction substitution }} (see (\ref{DL})):
\begin{equation}\label{dD}
    \frac{\partial}{\partial x^L} \rightarrow  D_L = \frac{\partial}{\partial x^L} - i q e^{iM x_5}A_L(x,x_5)
\end{equation}
and to take $x_5 = 0$.

2. Add to the obtained expression the action integral of the
electromagnetic field (\ref{5SA}) after setting  $x_5 = 0$.

The total action integral remains invariant under simultaneous gauge
transformations (\ref{gt1}) \footnote{ This relation must be taken on the
plane $x_5 = 0$.} and (\ref{ggcsf}).

A similar algorithm will be applied also in the new version of the
Standard Model.

If the neutral vector field has a mass $m$ (the so called Proca field) in
our approach it is described by the action (compare with (\ref{5SA})):
\begin{equation}\label{AProca}
   \begin{array}{c}
      S_0(M) = 2\pi M
  \int \varepsilon(p_5) \delta(p_L p^L - M^2)d\,^5p \times\\ \\

  \times\left\{2 M (p_5 - M\cos\mu
       )|A_n(p, p_5)|^2 - 2 p_n \overline{A}_n(p, p_5)A_5(p,
       p_5)\right\}-\\ \\

        - \left\{2 p_n \overline{A}_5(p, p_5)A_n(p, p_5) + 2(p_5 + M\cos\mu)|A_5(p, p_5)|^2 \right\}, \\
\end{array}
\end{equation}
where:
\begin{equation}\label{Procahc}
\cos\mu = \sqrt{1 - \frac{m^2}{M^2}}, \;\;A^+_n(p, p_5) = A_n(p, p_5),
\;\;A^+_5(p, p_5) = - A_5( - p, p_5).
\end{equation}
In configurational space, using (\ref{5FTA}) and introducing the notations
(compare with (\ref{phix} and \ref{chix}):
\begin{equation}\label{ALXL}
    \begin{array}{c}
      A_L(x) = A_L(x, 0) \\ \\
      X_L(x) = \frac{i}{M}\frac{\partial A_L(x, 0) }{\partial x^5}, \\
    \end{array}
\end{equation}
we obtain from (\ref{AProca}):
\begin{equation}\label{AProca}
   \begin{array}{c}
      S_0(M) = \int L_0 (x, M) d^4 x = \int  d^4 x L^{(0)}_{Proca} (x) + \\ \\
+ \frac{1}{2}\int  d^4 x \left[\frac{\partial A_k(x)}{\partial x_k} - iM
(A_5(x) + X_5(x))\right]^2 +\\ \\
+ \frac{M^2}{2}\int  d^4 x\left[ X_k(x)- \cos\mu
A_k(x)+\frac{i}{M}\frac{\partial A_5(x)}{\partial x_5} \right]^2 -
         \\ \\
         -2i M \sin^2\frac{\mu}{2}\int  d^4 x A_k(x)\frac{\partial
A_5(x)}{\partial x_5} - 2 M^2 \sin^2\frac{\mu}{2}\int  d^4 x A_5(x)X_5(x),
\end{array}
\end{equation}
where:
\begin{equation}\label{Lproca}
    L^{(0)}_{Proca} (x) = \frac{1}{4} F^2_{kl}(x) + \frac{m^2}{2}A^2_n(x).
\end{equation}

If in our version of the Euclidean scalar electrodynamics, we shortly
discussed above, a spontaneous breaking of the gauge $U(1)$ - symmetry is
realized, using for this a boson potential of the type (compare with
(\ref{Ltot1}):
\begin{equation}\label{Uintsed}
    \frac{\lambda^2}{4}\left(|\varphi(x)|^2 + |\chi(x)|^2 - v^2\right),
\end{equation}
then in result the electromagnetic field obtains a mass \footnote{ Let us
notice, that this quantity does not depend of M an coincides with the
standard expression (compare with (\ref{Ltot1})).} $m = q v$, i.e.
transforms to a Proca field. The quadratic in the vector field part of the
total Lagrangian will include $L^{(0)}_{Proca} (x)$ (see (\ref{Lproca})),
and also the terms:
\begin{equation}\label{quadrtA}
\frac{1}{2}\left[\frac{\partial A_k(x)}{\partial x_k} - iM (A_5(x) +
X_5(x))\right]^2 + \frac{1}{2}\left[ X_k(x)-
A_k(x)+\frac{i}{M}\frac{\partial A_5(x)}{\partial x_5} \right]^2
\end{equation}
The difference between (\ref{quadrtA}) and the last four terms in the
Lagrangian $L_0(x, M)$ from (\ref{AProca}) is not essential. It disappears
after a shift of the auxiliary field variable $X_k(x)$ in (\ref{AProca}):
\begin{equation}\label{shiftX}
    X_k \rightarrow X_k - 2\sin^2\frac{\mu}{2}A_k
\end{equation}
and  fixing the gauge  \footnote{Let us emphasize that $L^{(0)}_{Proca}
(x)$ does not depend on $A_5(x)$.}$A_5(x) = 0$.

In the theory of the Yang-Mills field in de Sitter p-space (\ref{ds})
\cite{CDKMNCG} the  mechanism of generating mass of the vector particles
as a result of spontaneous breaking of the symmetry is qualitatively the
same as in the just discussed Abelean case. In particular, in the
framework of  Maximal Mass Model, which will be developed in detail in a
separate publication (see section 1.), it turns out that the masses  of
$W^\pm$ and $Z^0$ are the known  SM expressions in terms of the coupling
constants and the Higgs vacuum expectation value and  do not contain $M$.

At the end of this section we would like to explain why we prefer to
develop our approach in Euclidean terms and pass from anti de Sitter
p-space (\ref{ads}) to de Sitter p-space (\ref{ds}).

 Let us apply to
(\ref{dads}) 5-dimensional Fourier transform (compare with (\ref{FT2})):
\begin{equation}\label{5dFTads}
    \varphi(x, x_5) \equiv \frac{2M}{(2\pi)^{3/2}}\int e^{-i p_0 x_0 + \emph{\textbf{p}} \emph{\textbf{x}} -
  i p_5 x_5   }\delta(p^2_0 - \emph{\textbf{p}}^2 + p^2_5 - M^2)\varphi(p,
 p_5) d^5\; p.
\end{equation}
From here we find (compare with (\ref{phix}) and (\ref{chix})):
\begin{equation}\label{phichiads}
    \begin{array}{c}
       \varphi(x, 0) \equiv \varphi(x) = \frac{M}{(2\pi)^{3/2}}\int_{p^2 \leq M^2} e^{-i p x} d^4 p \frac{\varphi(p, |p_5|)
        + \varphi(p, -|p_5|)}{|p_5|} \\ \\
    \frac{i}{M} \frac{\partial\varphi(x, 0)}{\partial x_5} \equiv \chi(x) =
    \frac{1}{(2\pi)^{3/2}}\int_{p^2 \leq M^2} e^{-i p x} d^4 p \left[\varphi(p,
    |p_5|) -
         \varphi(p, -|p_5|)\right].\\
     \end{array}
\end{equation}
The principal difference of these expressions in comparison with
(\ref{phix}) and (\ref{chix}) is that in (\ref{phichiads}) there is a
limitation on the integration region: $p_0^2 - \emph{\textbf{p}}^2 \leq
M^2$. This fact sharply restricts the class of functions $\varphi(x)$ and
$\chi(x)$ and does not allow, in particular, to construct from them local
Lagrangians or to apply to them local gauge transformations. Rigorously
speaking eqs. (\ref{phichiads}) can not be treated ( without special
reservations ) as Cauchy data for the "ultra-hyperbolic" equation:
\begin{equation}\label{pentads}
    \left( \frac{\partial^2}{\partial x_0^2} + \frac{ \partial^2}{\partial x_5^2} -
    \frac{\partial^2}{\partial \emph{\textbf{x}}^2}+ M^2\right)\varphi(x, x_5)
    = 0,
\end{equation}
which (\ref{5dFTads}) satisfies. In mathematical physics there are
developed methods, which allow one to use partial differential equations
of ultra-hyperbolic type with Cauchy initial data. In technical plan we
consider this as more complicated procedure, than to work in the framework
of Euclidean QFT. Moreover, thanks to the locality of the Euclidean
formulation, coming back to the relativistic description is not a problem.

\section{De Sitter fermion fields}

As far as the new QFT is elaborated on the basis of de Sitter momentum
space (\ref{ds}) it is natural to suppose that in the developed approach
the fermion fields $\psi_\alpha(p, p_5)$ have to be de Sitter spinors, i.e
to transform under the four dimensional representation of the group SO(4,
1). Further on we shall use the following $\gamma$ - matrix basis
$(\gamma^4 = i\gamma^0)$:
\begin{equation}\label{gammam}
   \begin{array}{c}
     \gamma^L = (\gamma^1, \gamma^2, \gamma^3, \gamma^4, \gamma^5) \\ \\
     \left\{\gamma^L, \gamma^M\right\}  = 2 g^{LM},\\ \\
g^{LM} = \left(%
\begin{array}{ccccc}
  -1 & 0 & 0 & 0 & 0 \\
  0 & -1 & 0 & 0 & 0\\
  0 & 0 & -1 & 0 & 0 \\
  0 & 0 & 0 & -1 & 0 \\
  0 & 0 & 0 & 0 & 1 \\
\end{array}%
\right).
   \end{array}
\end{equation}
Obviously we have:
\begin{equation}\label{Dir}
\begin{array}{c}
   M^2 - p_L p^L = M^2 + p_n^2 - p_5^2 = ( M - p_L \gamma^L )( M + p_L \gamma^L
    ) = \\ \\
 = ( M + p^n \gamma^n - p^5 \gamma^5)(M - p^n \gamma^n + p^5
    \gamma^5).\\
\end{array}
\end{equation}
In the "flat limit" $M\rightarrow\infty$ the quantities $\psi_\alpha(p,
p_5)$ become Euclidean spinor fields, which are used in the construction
of different versions of Euclidean QFT for fermions.

It is clear that the relations (\ref{FT2}) - (\ref{chix}) , which we
considered in the theory of boson fields, exist also in its fermion
version. Let us write some of them without comments:
\begin{equation}\label{5FTF}
\psi(x, x_5) = \frac{2M}{(2\pi)^{3/2}}\int e^{-i p_K x^K}\delta(p_L p^L -
M^2)\psi(p,
 p_5) d^5\; p ,
\end{equation}
\begin{equation}\label{pentF}
    \left(\frac{\partial^2}{\partial x_5^2 } - \square + M^2\right)\psi(x, x_5) =
    0,
\end{equation}

\begin{equation}\label{psix}
\begin{array}{c}
  \psi(x, 0)\equiv \psi(x) = \frac{2M}{(2\pi)^{3/2}}\int e^{i p_n x^n} d^4\,p
  \frac{\psi_1(p) + \psi_2(p)}{|p_5|} = \\ \\
  = \frac{1}{(2\pi)^{3/2}}\int e^{i p_n x^n} \psi(p) d^4\,p\\
\end{array}
\end{equation}
\begin{equation}\label{chixF}
\begin{array}{c}
  \frac{i}{M}\frac{\partial\psi(x, 0)}{\partial x_5} \equiv \chi(x)=
\frac{1}{(2\pi)^{3/2}}\int e^{i p_n x^n} d^4 p
  \left[\psi_1(p) - \psi_2(p)\right] =  \\ \\
  = \frac{1}{(2\pi)^{3/2}}\int e^{i p_n x^n} \chi(p) d^4\,p
.\\
\end{array}
\end{equation}
The next step is the construction of the action integral for the fermion
field $\psi_\alpha(p, p_5)$. Here we will not follow our work
\cite{CDKMNCD}, where this problem has been solved in the spirit of the
Schwinger's approach \cite{Schw} with the use of 8-component real spinors
and preserving the reality of the action. Now we shall follow the
formulation of Osterwalder and Schrader \cite{OS7375} and write the
Euclidean fermion Lagrangian in the form:
\begin{equation}\label{EucFL}
\begin{array}{c}
  L_E(x)= \overline{\zeta}_E(x)\left(- i \gamma_n\frac{\partial}{\partial x^n} +
    m\right)\psi_E(x), \\ \\
  \left\{\gamma^n, \gamma^m\right\} = -2
    \delta^{nm}\;\;\; ( m,n = 1, 2, 3, 4).\\
\end{array}
\end{equation}
Here the spinor fields $\overline{\zeta}_E(x) = \zeta^+_E (x)\gamma ^4  $
and $\psi_E(x)$  are independent Grassmann variables, which are not
connected between themselves by Hermitian or complex conjugation.
Correspondingly the action is not Hermitian. The Osterwalder and Schrader
approach has been widely discussed in the literature
\cite{OS7375}\footnote{ By the way in the paper \cite{NW} the so called
Wick rotation is interpreted in terms of 5-dimensional space.} and here we
shall not go into details. It is easy  to convince oneself, that the
expression $2M(p_5 - M \cos \mu)$, which in our approach substitutes (see
eq.(\ref{S0M})) the Euclidean Klein-Gordon operator $p^2_n + m^2$ may be
represented as:
\begin{equation}\label{Dir3}
\begin{array}{c}
  2M(p_5 - M \cos \mu) =  \\ \\
  = \left[p_n\gamma^n - (p_5 -M)\gamma^5 + 2M sin \frac{\mu}{2}\right]
   \left[- p_n\gamma^n + (p_5 -M)\gamma^5 + 2M sin \frac{\mu}{2}\right]\\
\end{array}
\end{equation}
In the Euclidean  approximation (\ref{FlatLp}) the relation (\ref{Dir3})
takes the form:
\begin{equation}\label{fldir}
    p^2_n + m^2 = \left( p_n\gamma^n + m\right)\left(- p_n\gamma^n +
    m\right).
\end{equation}
Therefore, we may use the expression:
\begin{equation}\label{Dirop}
    \mathcal{D}(p, p_5) \equiv p_n\gamma^n - (p_5 -M)\gamma^5 + 2M sin \frac{\mu}{2}
\end{equation}
like the new Dirac operator.

As a result we come to an expression for the action of the Fermion field
in the de Sitter momentum space:
\begin{equation}\label{FErA}
    \begin{array}{c}
       S_0(M) = 2\pi M \int \varepsilon(p_5) \delta(p_L p^L - M^2)d^5 p\times  \\ \\
      \times \left[\;\;\overline{\zeta}(p, p_5)(p_n\gamma^n - (p_5 -M)\gamma^5
      + 2M\sin\frac{\mu}{2})\psi(p, p_5)\;\right], \\
\end{array}
\end{equation}
In the integral (\ref{FErA}) it is possible to pass to the field
variables:
\begin{equation}\label{fvar}
    \begin{array}{c}
      \psi(p) = \frac{M}{|p_5|} (\psi(p, |p_5|) + \psi(p, -|p_5|)) \equiv M \frac{\psi_1(p) + \psi_2(p)}{|p_5|}\\ \\
      \chi (p) = \psi_1(p) - \psi_2(p)\\ \\
      \overline{\zeta}(p) = M \frac{\overline{\zeta}_1(p) + \overline{\zeta}_2(p)}{|p_5|} \\ \\
      \overline{\xi}(p) = \overline{\zeta}_1(p) - \overline{\zeta}_2(p),\\
    \end{array}
\end{equation}
which are the Fourier amplitudes of the local fields $\psi(x), \chi(x),
\overline{\zeta}(x)$ and $\overline{\xi}$ (x) (compare with (\ref{psix})
and (\ref{chixF})). As  result we get:
\begin{equation}\label{S0D}
\begin{array}{c}
   S_0^{\mathcal{D}} = - \pi\int d^4 p \left(M + \frac{p^2_n}{M}\right)\overline{\zeta}(p)\gamma^5
   \psi(p)+\\ \\
  + \pi\int d^4 p\overline{\zeta}(p)\left(\slashed{p} + M\gamma^5 + 2M sin \frac{\mu}{2}\right)\chi (p)
  +\\ \\
  + \pi\int d^4 p \overline{\xi(p)}\left(\slashed{p} + M\gamma^5 + 2M sin \frac{\mu}{2}\right)\psi (p) -
  \\ \\
 - \pi\int d^4 p M \overline{\xi(p)}\gamma^5 \chi (p)  \\ \\
\end{array}
\end{equation}
In configuration space we shall have correspondingly:
\begin{equation}\label{S0Dx}
    \begin{array}{c}
      S_0^{\mathcal{D}} = \int L_0^\mathcal{D}(x, M) d^4x =  \\ \\
      = \frac{1}{2} \int d^4 x \overline{\zeta}(x)\left(\frac{\square}{M^2} - 1\right)\gamma^5 \psi(x)
      +\\ \\
      + \frac{1}{2} \int d^4 x \overline{\zeta}(x)\left( i \gamma^n \frac{\partial}{\partial x^n} +
      M\gamma^5 +
      2M sin \frac{\mu}{2}\right)\chi (x) + \\ \\
       + \frac{1}{2} \int d^4 x \overline{\xi(p)}\left( i \gamma^n \frac{\partial}{\partial x^n} +
       M\gamma^5 +
      2M sin \frac{\mu}{2}\right)\psi (x) -  \\ \\
      - \frac{1}{2} \int d^4 x \overline{\xi(x)}\gamma^5 \chi (x).
    \end{array}
\end{equation}
Hence,  the modified Dirac Lagrangian $L_0^\mathcal{D}(x, M) $ is a local
function of the spinor field variables  $\psi(x), \chi(x),
\overline{\zeta}(x)$ and $\overline{\xi}$ (x). Here there is an obvious
analogy with the boson case (compare with (\ref{S0M}) and (\ref{AProca})).

However the fermion Lagrangian $L_0^\mathcal{D}(x, M) $ may be represented
in an other  form, if one use the relations (\ref{Dir}). Indeed let us
put:
\begin{equation}\label{prop}
\begin{array}{c}
   \frac{1}{2M}(M - p_K \gamma^K) \psi(p, p_5) \equiv \Pi_L \psi(p, p_5) \equiv \psi_L(p, p_5)\\ \\
  \frac{1}{2M}(M + p_K \gamma^K) \psi(p, p_5) \equiv \Pi_R \psi(p, p_5) \equiv \psi_R(p, p_5) \\
\end{array}
\end{equation}
Because of (\ref{ds}) the operators $\Pi_L$ and $\Pi_R$ are projectors:
\begin{equation}\label{progpr}
    \begin{array}{c}
       \Pi_L + \Pi_R = 1,\\ \\
       \Pi^2_L = \Pi_L \;\;\;\;\;\;\Pi^2_R = \Pi_R,\\ \\
       \Pi_L \Pi_R = \Pi_R \Pi_L = 0. \\
     \end{array}
\end{equation}

From other point of view they are the 5- analogue of the Dirac operator,
and the fields $\psi_L(p, p_5)$ and $\psi_R(p, p_5)$ obviously satisfy the
corresponding 5-dimensional Dirac equations:
\begin{equation}\label{5ddir}
    \begin{array}{c}
       (M + p_K \gamma^K) \psi_L(p, p_5) = 0, \\ \\
      (M - p_K \gamma^K) \psi_R(p, p_5) = 0. \\
    \end{array}
\end{equation}
Therefore,  in this way the fermion field $\psi(p, p_5)$, given in the de
Sitter momentum space (\ref{ds}), may be presented as a sum of two fields
$\psi_L(p, p_5)$ and $\psi_R(p, p_5)$:
\begin{equation}\label{SLR}
    \psi(p, p_5) = \psi_L(p, p_5) + \psi_R(p, p_5),
\end{equation}
which obey the 5-dimensional Dirac equations (\ref{5ddir}). Obviously the
decomposition (\ref{SLR}) is \emph{\textbf{de Sitter invariant
procedure}}.

It is easy to verify that in the flat limit (\ref{FlatLp}):
\begin{equation}\label{flpi}
   \Pi_{L, R} = \frac{1 \mp \gamma^5}{2},
\end{equation}
This is the reason that we consider the fields $\psi_L(p, p_5)$ and
$\psi_R(p, p_5)$ as the "\emph{\textbf{chiral}}" components in the
developed approach \cite{kadFursSpin}. The new operator of chirality
$\frac{p_L \gamma^L}{M}$, similarly to its "\emph{\textbf{flat
counterpart}}" has eigenvalues equal to $\pm1$, but \emph{\textbf{depends
on the energy and momentum}}. The last circumstance, as we hope, should be
revealed experimentally (see section 4).

It is worthwhile to pass in (\ref{5ddir}) to configurational
representation. Applying (\ref{5FTF}) we get :
\begin{equation}\label{LRx}
    \begin{array}{c}
    \psi_L(x, x_5) = \frac{1}{2} \left( 1 - \frac{i\gamma^n}{M}\frac{\partial}
    {\partial x^n} - \frac{i\gamma^5}{M}\frac{\partial}
    {\partial x^5 }\right) \psi(x, x_5) \\ \\
    \psi_R(x, x_5) = \frac{1}{2} \left( 1 + \frac{i\gamma^n}{M}\frac{\partial}
    {\partial x^n} + \frac{i\gamma^5}{M}\frac{\partial}
    {\partial x^5} \right) \psi(x, x_5)  \\
    \end{array}
\end{equation}
Setting in (\ref{LRx}) $x_5 = 0$ and taking into account (\ref{psix}) and
(\ref{chixF}) we shall have:
\begin{equation}\label{LRx0}
    \begin{array}{c}
      \psi_L(x, 0) \equiv \psi_{(L)}(x) = \frac{1}{2} \left( 1 - \frac{i\gamma^n}{M}\frac{\partial}
    {\partial x^n}\right)\psi(x) - \frac{\gamma^5}{2}\chi(x),\\ \\
      \psi_R(x, 0) \equiv \psi_{(R)}(x) = \frac{1}{2} \left( 1 + \frac{i\gamma^n}{M}\frac{\partial}
    {\partial x^n}\right)\psi(x) + \frac{\gamma^5}{2}\chi(x). \\
    \end{array}
\end{equation}
As far as the field $ \psi(x, x_5)$ obeys equation (\ref{pent}) the
relations, which we obtained for the scalar field in the "flat"
approximation and in particular (\ref{phi-chifl}),  may be applied to it.
Taking this into account, we find, that in this approximation the
equalities (\ref{LRx0}) become:
\begin{equation}\label{LRflat}
    \begin{array}{c}
       \psi_{(L)}(x) = \frac{1}{2}(1 - \gamma_5)\psi(x) - \frac{i\gamma^n}{2 M}\frac{\partial}
    {\partial x^n}\psi(x) + \frac{\gamma^5}{2}(\psi(x)- \chi(x)) \simeq \\ \\
       \simeq \frac{1}{2}(1 - \gamma_5)\psi(x) - \frac{i\gamma^n}{2 M}\frac{\partial}
    {\partial x^n}\psi(x) + \frac{\gamma^5}{4M^2}\square\psi(x), \\ \\
\psi_{(R)}(x) \simeq \frac{1}{2}(1 + \gamma_5)\psi(x) + \frac{i\gamma^n}{2
M}\frac{\partial}
    {\partial x^n}\psi(x) - \frac{\gamma^5}{4M^2}\square\psi(x).
     \end{array}
\end{equation}
Representation, analogous to (\ref{SLR}), may be introduced for the field
$\overline{\zeta}(p, p_5)$, appearing in (\ref{FErA}):
\begin{equation}\label{zetaRL}
    \overline{\zeta}(p, p_5) = \overline{\zeta}_L(p, p_5) + \overline{\zeta}_R(p, p_5),
\end{equation}
where
\begin{equation}\label{zetaRL1}
    \begin{array}{c}
      \overline{\zeta}_L(p, p_5) = \overline{\zeta}(p, p_5)\Pi_R, \\ \\
      \overline{\zeta}_R(p, p_5) = \overline{\zeta}(p, p_5)\Pi_L. \\ \\
    \end{array}
\end{equation}
Further it is not difficult to obtain relations, similar to (\ref{LRx})
and (\ref{LRflat}) for the fields $\overline{\zeta}_L(x)$ and
$\overline{\zeta}_R(x)$:
\begin{equation}\label{zetax}
    \begin{array}{c}
      \overline{\zeta}_{(L)}(x) = \frac{1}{2} \overline{\zeta}(x)  + \frac{i}{M}\frac{\partial \overline{\zeta}(x)}
    {\partial x^n}\gamma^n + \overline{\xi}(x)\frac{\gamma^5}{2}, \\ \\
      \overline{\zeta}_{(R)}(x) = \frac{1}{2} \overline{\zeta}(x)  - \frac{i}{M}\frac{\partial \overline{\zeta}(x)}
    {\partial x^n}\gamma^n - \overline{\xi}(x)\frac{\gamma^5}{2},\\
    \end{array}
\end{equation}

\begin{equation}\label{zetaxflat}
\begin{array}{c}
      \overline{\zeta}_{(L)} \simeq \overline{\zeta}(x) \frac{1}{2}(1 + \gamma_5) + \frac{i}{M}\frac{\partial}
    {\partial x^n}\overline{\zeta}(x) \frac{\gamma^n}{2
} - \frac{\square}{2 M^2}\overline{\zeta}(x) \frac{\gamma^5}{2}, \\ \\
      \overline{\zeta}_{(R)} \simeq\overline{\zeta}(x) \frac{1}{2}(1 - \gamma_5) - \frac{i}{M}\frac{\partial}
    {\partial x^n}\overline{\zeta}(x) \frac{\gamma^n}{2
} + \frac{\square}{2 M^2}\overline{\zeta}(x) \frac{\gamma^5}{2}. \\
    \end{array}
\end{equation}
Now, substituting (\ref{LRx0}) and (\ref{zetax}) in the action integral
(\ref{S0Dx}) we may pass to new variables $\psi_{(L)}(x), \psi_{(R)}(x),
\overline{\zeta}_L(x)$ and $\overline{\zeta}_R(x)$:
\begin{equation}\label{SODx}
    \begin{array}{c}
      S_0^{\mathcal{D}} = \int L_0^\mathcal{D}(x, M) d^4x =  \\  \\
      = \int d^4 x\left[\overline{\zeta}_{(L)}i \gamma^n \frac{\partial}{\partial x^n} \psi_{(L)}(x)
       + \overline{\zeta}_{(R)}i \gamma^n \frac{\partial}{\partial x^n} \psi_{(R)}(x)\right] +\\ \\
     + \int d^4 x\overline{\zeta}_{(L)}\left[i \gamma^n \frac{\partial}{\partial x^n} + M (1 - \gamma^5)\right]\psi_{(R)}(x) + \\ \\
     + \int d^4 x \overline{\zeta}_{(R)}\left[i \gamma^n \frac{\partial}{\partial x^n} - M (1 + \gamma^5)\right]\psi_{(L)}(x) + \\ \\
      + 2 M \sin \frac{\mu}{2}\int d^4 x\left[ \overline{\zeta}_{(L)}(x) \gamma^5 \psi_{(R)}(x) - \overline{\zeta}_{(R)}(x)
       \gamma^5 \psi_{(L)}(x)\right]\\ \\
    \end{array}
\end{equation}
The obtained expression is the basis for constructing gauge theory of
interacting fermion field. This topic we shall discuss shortly in the next
section. Concluding this part we would like to make one important remark
\cite{CDKMNCD}.

The point is that for the quantity $2M(p^5 - M \cos \mu)$, which
substituted in our approach the Euclidean Klein-Gordon  operator together
with (\ref{Dir3}) there exists \emph{\textbf{one more decomposition to
matrix factors}}:
\begin{equation}\label{exdir}
    \begin{array}{c}
    2M(p^5 - M \cos \mu) =  \\ \\
      = (p_n\gamma^n - \gamma^5 (p^5 + M) + 2M\cos\frac{\mu}{2})(p_n\gamma^n + \gamma^5 (p^5 + M) - 2M\cos\frac{\mu}{2}) \\
    \end{array}
\end{equation}
Therefore, if our approach is considered to be realistic, it may be
assumed that in Nature exists some \emph{\textbf{exotic }} fermion field,
whose free action integral has the form:
\begin{equation}\label{S0exotic}
 \begin{array}{c}
       S_0^{(exotic)}(M) = 2\pi M \int \varepsilon(p_5) \delta(p_L p^L - M^2)d^5 p \;\times  \\ \\
      \times \left\{\;\;\overline{\xi}_{exotic}(p, p_5)\left[p_n\gamma^n - (p_5 + M)\gamma^5
      + 2M \cos\frac{\mu}{2}\right]\psi_{exotic}(p, p_5)\;\right\}\\
\end{array}
\end{equation}

Applying the above developed procedure it is easy to obtain
$S_0^{(exotic)}(M)$ in a form analogous to (\ref{SODx}). However, in
contrast to $ S_0^{\mathcal{D}}$ this quantity does not have a limit at
$M\rightarrow\infty$, which justifies the chosen by us name for this
field. The polarization properties of the exotic field, evidently, differ
sharply from standard ones.

We would like to conjecture that the quanta of the exotic fermion field
have a direct relation to the structure of the "dark matter."

\section{The new geometrical approach to the Standard Model}

To the complete formulation of the Standard Model, consistent with the
principle of maximal mass (\ref{Mfund}) and its geometrical realization in
terms of de Sitter momentum space \footnote{Let us recall that namely this
\emph{\textbf{geometrized}} SM we call in advance the Maximal Mass Model.}
(\ref{ds}) we shall devote a separate paper. Now we intend to make only
several remarks, important for the understanding of our general strategy:

1. $SU_L(2)\bigotimes U_Y(1)$ - \emph{\textbf{symmetry}}

The gauge $SU_L(2)\bigotimes U_Y(1)$ - symmetry is one of the most
important elements of the SM, which guaranteed its success. This is why it
should be assumed as necessary to apply it also in our approach, taking
into account our new definition of the chiral fields. However, in the new
fermion Lagrangian $ L_0^{\mathcal{D}}$ (see (\ref{SODx})) even for $m=0$
there are crossed terms:
\begin{equation}\label{ct}
   \begin{array}{c}
    \overline{\zeta}_{(L)}\left[i \gamma^n \frac{\partial}{\partial x^n} + M (1 - \gamma^5)\right]\psi_{(R)}(x) +   \\ \\
    + \overline{\zeta}_{(R)}\left[i \gamma^n \frac{\partial}{\partial x^n} - M (1 + \gamma^5)\right]\psi_{(L)}(x)\\
   \end{array},
\end{equation}
which, at first glance are insurmountable obstacle for the use of the
group $SU_L(2)\bigotimes U_Y(1)$. The solution of  this difficulty is to
make the expression (\ref{ct}) invariant form with the help of the Higgs
field. In this way, considering as before the Higgs boson to be a
$SU_L(2)$-doublet, introducing the doublet structure for the $L$-component
of the fermion field and passing to covariant derivatives with the rules
of the SM, we may write (\ref{ct}) in the form:
\begin{equation}\label{ctMMM}
\begin{array}{c}
  \frac{1}{v} \left(\overline{\zeta}_{(L)}. H(x)\right)\left[i\gamma^n D^R_n + M (1 - \gamma^5)\right] \psi_{(R)}(x) + \\ \\
  + \frac{1}{v}\overline{\zeta}_{(R)}\left\{H^+(x).\left[i\gamma^n D^L_n - M (1 + \gamma^5)\right]\psi_{(L)}(x)\right\} + conj., \\
\end{array}
\end{equation}
where $H(x)$ is the SM Higgs doublet and $D^R$ and $D^L$ are the SM
covariant derivatives. After the Higgs mechanism is switched on from
(\ref{ctMMM}) separate our cross terms (\ref{ct}) and appear terms with
interactions, which are not present in the SM. Together with the
corrections, caused by the difference between the new and old definitions
of chirality (see (\ref{LRflat}) and (\ref{zetaxflat})) they may be the
ground for predictions, which may be verified experimentally.

2.\emph{\textbf{ Chirality}}

In the SM to the boson fields it is prescribed to transform as
representations of the group $SU_L(2)$, which for the vector fields is
three-dimensional and two-dimensional for the Higgs scalar. Naively
reasoning one may ask himself how the mentioned bosons should know about
the existence of the $4\times4$ matrix $\gamma^5$, one of the eigenvalues
of which, corresponds to the index $L$ ? In our approach all fields, boson
and fermion, are given in de Sitter p-space on an equal footing, with the
only difference that the boson fields obey the 5-equation of Klein-Gordon
(see (\ref{pent}) and (\ref{pentA})), and the fermion 5-equations of Dirac
(\ref{5ddir}). There is nothing strange in it that the field
$\psi_{(L)}(x)$ and the Higgs scalar $\varphi$ simultaneously have a
doublet structure in respect to the $SU_L(2)$-symmetry. This has already
happened in the old isospin symmetry. Let us recall the nucleon doublet
and the K-meson doublet.

The new geometrical concept of chirality allows us to think that the
discovered fifty years ago parity violation in weak interactions was a
manifestation of the de Sitter nature of momentum 4-space.

3.\emph{\textbf{ Higgs mechanism}}

This important element of the SM, as we see already now is conserved in
the generalized SM without considerable changes. The role of the
spontaneous symmetry breaking mechanism in the formation of the
fundamental mass $M$ has been studied on a simple example in section 2.

\section{Concluding remarks}

Concluding this article, we would like to pay attention to one peculiarity
of the developed here approach. All fields, independent of their spins,
charges, masses etc. satisfy the free 5-equation  of hyperbolic type, and
the role of "time" is played by the coordinate "$x_5$". The interaction
between the fields is realized on the level of the Cauchy data, given on
the plane $x_5 = 0$, i.e. in the four-dimensional (Euclidean) world. The
right of such a "\emph{\textbf{free gliding}}" in the 5-space have only
the elementary particles, described by local fields and with masses,
obeying the limitation $m \leq M$.

 {\bf Acknowledgements.} V. G. K. and M. D. M. would like to thank Luis Alvarez-Gaum\'{e}
 and the PH - TH division of CERN, where major part of this work was
done, for the kind hospitality and stimulating atmosphere. Both authors
would like to thank warmly Alvaro De R{\'{u}}jula for fruitful and highly
useful discussions. The authors would like to thank also M. V. Chizhov and
E. R. Popitz
 for discussions. This work has been supported in part by
the Bulgarian National Science Fund under the contract Ph-09-05 (M.D.M.)
and by the Russian Foundation for Basic Research (Grant No. 05-02-16535a)
and the Program for Supporting Leading Scientific Schools (Grant No.
NSh-5332.2006.2) (V. G. K. and V. N. R.).

\end{document}